\documentclass[aps,prb,twocolumn,superscriptaddress,amsmath,showpacs]{revtex4}
\usepackage{graphicx}

\begin{document}
\title{The nature of an enhanced ferroelectric phase transition temperature
in perovskite-based solid solutions}
\author{V.A.Stephanovich}
\homepage{http://cs.uni.opole.pl/~stef}
\email{stef@math.uni.opole.pl} \affiliation {Opole University,
Institute of Mathematics and Informatics, Opole, 45-052, Poland}
\author{M.D.Glinchuk}
\email{glin@materials.kiev.ua} \affiliation{Institute for Problems
of Materials Science, National Academy of Science of Ukraine,\\
Krjijanovskogo 3, 03142 Kiev, Ukraine}
\author{C.A.Randall}
\affiliation{The Pennsylvania State University, 144 Materials
Research Laboratory Building, \\University Park, PA 16802, USA}
\date{\today }
\pacs{77.80.Bh,77.84.Lf,77.84.Dy}
\begin{abstract}

We explain the phenomena of ferroelectric phase transition
temperature $T_c$ enhancement beyond the end members in perovskite
solid solutions like BiMeO$_3$-PbTiO$_3$ (Me=Sc, In, etc.) is
related to nonlinear and spatial correlation effects. The
explanation is based on the calculation of $T_c$ in the framework
of our random field theory with additional account for nonlinear
effects in the above substances. We show that the maximum of $T_c$
for certain PbTiO$_3$ content takes place when coefficient of
nonlinearity is positive, the value of this coefficient is found
from best fit between theory and experiment. This nonlinearity
coefficient is the only adjustable parameter of the theory. We
show that enhancement of positive nonlinearity coefficients
enhances greatly the $T_c$ maximum over its value for end members.

The theory lays the foundation to calculate not only $T_c$ for
above solid solutions but virtually any equilibrium and/or
nonequilibrium thermodynamic characteristics such as static and
dynamic dielectric susceptibility, specific heat etc as a
functions of PbTiO$_3$ content, temperature, electric field and
other external parameters.
\end{abstract}
\maketitle

\section{Introduction}

Over the last few years considerable effort has been spared to
synthesize the dielectric materials with controllable properties
for many technical applications. The most promising substances are
believed to be the compound materials consisting of solid
solutions of different combinations of ferroelectrics with
different dielectric properties. One of the examples is perovskite
solid solutions like BiMeO$_3$-PbTiO$_3$ (Me=Sc, In, etc.), which
have high ferroelectric phase transition temperatures $T_c$ at the
morphotropic phase boundaries with enhancement beyond the end
members. Such materials can be used as the materials with
excellent high-temperature piezoelectric
properties\cite{rb1,rb2,rb3,rb4,rb5,rb6}.

A common feature of the above solid solutions is the existence of
numerous random fields sources due to substitutional disorder,
unavoidable impurities, vacancies in anion and cation sublattices
etc. These random fields play a crucial role in the properties of
disordered ferroelectric and magnetic materials (see e.g.
\cite{GS1,GS2,Koshe}) and above substances in particular. This
means that observable physical properties of above systems depend
strongly on the form of random fields distribution function.
Namely, the relation between width of the distribution function
(the dispersion of random fields) and its first moment (mean value
of random field) generates all observable features of the phase
diagram of a disordered dielectric and/or magnetic materials, i.e.
realization of ferroelectric (ferromagnetic in the case of
magnetic materials) phase, dipole (spin) glass phase and mixed
ferroglass phases. Also, so-called paraglass (Griffiths) phase may
occur in disordered dielectrics (see, e.g. \cite{SG}).

On the other hand, both ordered and disordered dielectrics have
intrinsic nonlinearities, consisting in, e.g., dielectric
hysteresis. The "interaction" between these nonlinearities and
random fields lead to their renormalization so that the
distribution function of random fields will also include a
nonlinear contribution of random fields. Such calculation had been
carried out in a different context in Ref.\cite{Glfarstef}.

The calculations in Ref.\cite{Glfarstef} incorporate a
self-consistent dependence of distribution function of random
fields on the third order nonlinearity coefficient $\alpha _3$. It
has been shown that when the nonlinear coefficient is sufficiently
large and positive, the results are strongly different from those
in linear case \cite{GS1}. In particular, for $\alpha _3>0$, the
phase transition temperature exceeds its mean field asymptotic
value, while for $\alpha _3<0$ the results are qualitatively the
same as in linear case. As we have found out earlier, this
phenomenon is due to the "generation of order by disorder" (or
more precisely, a specific positive feedback generated by positive
nonlinearity) taking place at $\alpha _3>0$.

\section{General formalism}
\subsection{The distribution function of random fields}
Here we briefly review the main facts about the shape of
distribution function of random fields with respect to nonlinear
effects, a more detailed discussion can be found, e.g. in Refs
\cite{GS1,Glfarstef} for disordered dielectrics and in Refs
\cite{SS,SS2} for disordered magnetic semiconductors.

The distribution function of random field $\vec E$ can be
represented in the form
\begin{equation}\label{re1}
f({\vec E})=\overline{\left\langle \delta \left({\vec E}-%
{\vec E}({\vec r_i})\right)\right\rangle }
\end{equation}
Here the bar denotes averaging over spatial configurations of
random fields sources (e.g. electric dipoles, "responsible" for
emergence of ferroelectricity in the above compounds), $\langle
...\rangle $ means the averaging over dipoles orientations, ${\vec
E}({\vec r_i})$ is the internal electric field induced by electric
dipoles and other sources in the observation point ${\vec r_i}$.
In a disordered ferroelectric this field already contains the
intrinsic nonlinearity and can be written in the form (see
\cite{Glfarstef} for details)
\begin{equation}\label{re2}
E_{\gamma }({\vec r_i})= {\cal E}_{\gamma}({\vec r_i})+
\sum\limits_{m=2}^{\infty}\alpha _m\prod_{j=1}^{m}{\cal
E}_{{\gamma}_j}({\vec r_i}),
\end{equation}
where $\gamma $=x,y,z so that ${\cal E}_{\gamma}$ denotes simply
the $\gamma $-component of vector ${\vec {\cal E}}$. Here ${\vec
{\cal E}}$ is an internal electric field induced by electric
dipoles, $\alpha _m$ is a coefficient of nonlinearity of $m$-th
order. Note that the first term in Eq. (\ref{re2}) can be
generalized to account for other possible sources of random fields
such as point charges, elastic dipoles etc
\cite{rb5,GS1,Glfarstef}.

The calculation of the distribution function (\ref{re1}) with
respect to (\ref{re2}) in the framework of statistical theory (see
\cite{sto,book} for details of this theory) for so-called
disordered Ising model (when the dipole has only two admissible
orientations) yields the following rigorous result for any
electric field component $E_\alpha \equiv E$ \cite{Glkond}
\begin{equation}\label{re3}
f(E)={\int }_{-\infty }^{\infty}f_1(E')\delta
\left(E-E'-\sum\limits_{m=2}^{\infty}\alpha _m{E'}^m\right)dE'
\end{equation}
where $f_1(E)$ is the distribution function, that takes into
account only the first linear term in (\ref{re2}) (distribution
function of the first order).

More detailed version of Eq. (\ref{re3}) reads
\begin{eqnarray}
&&f(E)={\int }_{-\infty }^{\infty }f_1(E')\times \nonumber \\
&&\times \delta \left(E-E' -\sum\limits_{m=2}^{\infty}\alpha _m{E'}^m\right)dE',\label{re3a}\\
&&f_1(E)=\frac 1{2\pi }\int_{-\infty }^{\infty }e^{iEt-nF(t)}dt,\label{re4}\\
&&F(t)=\int d^3r\left<\left(1-e^{-it{\cal E} ({\vec
r})}\right)\right>\label{re5}.
\end{eqnarray}
Here $n$ and ${\cal E} ({\vec r})$ are the concentration and
electric field of the dipoles. The Eq.(\ref{re4}) determines the
function $f_1(E)$, calculated earlier in \cite{GS1} for the case
of two-orientable electric dipoles. We note here, that in general
case (e.g. arbitrary interaction between above dipoles, their
arbitrary concentration etc, see Ref. \cite{GS1} and references
therein for discussion) function $f_1(E)$ has non-Gaussian form.
It can be shown that for our case of compound ferroelectrics it is
sufficient to use its Gaussian asymptotics, which reads
\begin{equation}\label{re6}
f_1(E)=\frac 1{2\sqrt{\pi nB}} \exp
\left[-\frac{\left(E-E_0L\right)^2}{4nB}\right].
\end{equation}
Here $L=\frac{\overline{\left< d^{*}\right\rangle}}{d^{*}}$ and
$E_0=4\pi (nd^{*2})/\varepsilon _0$ are the order parameter
(number of coherently oriented impurity electric dipoles or
dimensionless spontaneous polarization) and the mean value of
random field of electric dipoles (in the energy units), $%
d^{*}=\frac 13d\gamma (\varepsilon _0-1)$ is the effective
electric dipole moment, $\gamma $ and $\varepsilon _0$ are,
respectively, Lorentz factor and static dielectric permittivity of
the host lattice, $n$ is the concentration of electric dipoles.
Coefficient $B$ determines the width of the distribution function
(dispersion of random fields) and depends on host lattice
parameters like its correlation radius $r_c$, see \cite{GS1} for
details.
\subsection{The equation for long-range order parameter}
In our approach, an average value $\overline{A}$ of any physical
quantity can be represented in the form

\begin{equation}\label{re7}
\overline{A}=\int_{-\infty }^\infty f(E)A(E)dE,
\end{equation}
where $f(E)$ is determined by (\ref{re3a}) and $A(E)$ is the above
quantity for single random field constituent, averaged over its
internal degrees of freedom. In our case this is the average value
over the orientations of single electric dipole.

To calculate the $T_c$ for disordered ferroelectric compound we
need to calculate first the long-range order parameter $L$. In the
spirit of Eq. (\ref{re7}) we obtain following self-consistent
equation for this parameter
\begin{widetext}
\begin{equation}\label{re8}
L=\int_{-\infty }^\infty f_1(E)\tanh \left[ \left(E +\sum\limits
_{m=2}^{\infty}\alpha _mE ^m\right)/kT\right] dE.
\end{equation}
\end{widetext}
Here we use the fact that for two-orientable dipoles ($l_z=\pm 1,$
$l_x=l_y=0$) $A(E)=<l> =\tanh \left( \frac E{kT}\right)$, $E\equiv
E_z$.

The self-consistency of Eq. (\ref{re7}) is revealed by
substitution of Eq.(\ref{re7}) into it, which yields
\begin{widetext}
\begin{equation}\label{re9}
L=\frac 1{2\sqrt{\pi nB}}\int_{-\infty }^\infty  \exp
\left[-\frac{\left(E-E_0L\right)^2}{4nB}\right] \tanh \left[
\left(E +\sum\limits _{m=2}^{\infty}\alpha _mE ^m\right)/kT\right]
dE.
\end{equation}
\end{widetext}
In the linear case ($\alpha _m=0$) Eq.(\ref{re9}) transforms into
that derived in \cite{GS1}. It is seen that order parameter is
self-consistently expressed through itself and is the function of
temperature, dipoles concentrations and nonlinearity coefficients.

Further simplifications of Eq. (\ref{re9}) are possible on
symmetry grounds. Namely, for the lattice with the center of
inversion in paraelectric phase, the order parameter has to be an
odd function of electric field, i.e. $m$'s in Eq.(\ref{re9}) are
odd numbers. Conserving only the first nonlinear term in $\tanh $
argument, we obtain
\begin{widetext}
\begin{equation}\label{re10}
L=\frac 1{2\sqrt{\pi nB}}\int_{-\infty }^\infty  \exp
\left[-\frac{\left(E-E_0L\right)^2}{4nB}\right] \tanh \left[
\left(E +\alpha _3E ^3\right)/kT\right] dE.
\end{equation}
\end{widetext}
It is now instructive to consider the mean field limit of equation
(\ref{re10}). This limit corresponds to the case of an ordered
ferroelectric, where the distribution function of random fields
degenerates into a delta function $\delta (E -E_0L)$. Formally, in
our method this limit corresponds to $nr_c^3\to \infty $.
Substitution of this $\delta $-function into (\ref{re10}) gives
the desired mean field equation for order parameter

\begin{equation}\label{re10a}
L_{mf}=\tanh \left[ \frac{T_{cmf}}T(L_{mf}+\alpha
_0L_{mf}^3)\right]
\end{equation}
where $E_0=kT_{cmf}$, $T_{cmf}$ is transition temperature in a
mean field approximation (see below), $\alpha _0\equiv \alpha
_3E_0^2$ so that $\alpha _0$ is dimensionless. We now use the Eq.
(\ref{re10}) to derive the equation for ferroelectric phase
transition temperature beyond the mean field approximation.

\subsection{Ferroelectric phase transition temperature}
The phase transition temperature is defined as a temperature when
a nonzero order parameter $L$ appears. In other words, to get an
equation for $T_c$ from (\ref{re10}), we should put in it $L\to
0$. This is accomplished by noting that at small $L$
\begin{equation}\label{re11}
\exp\left[-\frac{\left(E-E_0L\right)^2}{4nB}\right] \approx
e^{-\frac{E^2}{4nB}}\left(1+\frac{EE_0L}{2nB}\right).
\end{equation}
Subsequent substitution of (\ref{re11}) into (\ref{re10}) yields
after some algebra
\begin{equation}\label{tauc}
\frac{\lambda }{\tau _c}\int\limits_{0}^{\infty}\frac{(1+3\alpha
_0x^2 )e^{-\frac{\pi }{4}x^2\lambda ^2}}{\cosh^2\frac{x+\alpha
_0x^3 }{\tau _c}}dx=1.
\end{equation}
Here we introduced the following dimensionless variables
\begin{eqnarray}
&&\lambda =\frac{E_0}{\pi nB}\equiv \sqrt{15nr_c^3},\nonumber \\
&&x=\frac{E}{E_0},\quad \tau _c=\frac{kT_c}{E_0}\equiv
\frac{T_c}{T_{cmf}}.\label{dim}
\end{eqnarray}
Equation (\ref{tauc}) is the main theoretical result of this work.
It predicts the existence of the critical concentration of dipoles
\[
n_{cr}r_c^3=\frac{\lambda _{cr}^2}{15},
\]
such that for $n<n_{cr}$, the long range order in the system would
never be realized. Thus, the critical concentration is determined
from the condition, that $\lambda = \lambda _{cr}$ at $\tau _c=0$.
Taking the limit $\tau _c\to 0$ \cite{ncr} in Eq. (\ref{tauc}), we
obtain $\lambda _{cr}=1$, which justifies the choice of
dimensionless parameter $\lambda $. It is seen that the critical
concentration does not depend on the coefficient of nonlinearity
and it is completely the same as in linear case (see \cite{GS1,
Glfarstef}).

\begin{figure}[th]
\vspace*{-5mm} \hspace*{-5mm} \centering{\
\includegraphics[width=8cm]{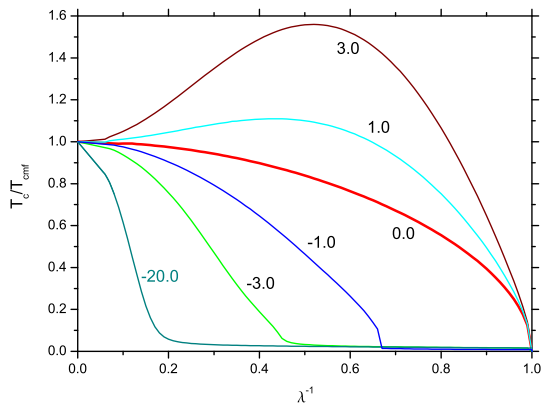}}
\caption{Theoretical dependence $\tau _c$$(\lambda ^{-1})$.
Numbers near curves denote magnitudes of non-linear coefficient,
$\alpha _0$.} \label{fig:te}
\end{figure}

 Now we demonstrate that in a mean field approximation
$\tau _{cmf}=1$, i.e. that $E_0=kT_{cmf}$. For that we notice that
this approximation corresponds to $\lambda \to \infty$ in
(\ref{tauc}), which after some transformations \cite{ncr} gives
$\tau _{cmf}=1$. This value also does not depend on the
coefficient of nonlinearity.

 The plot of dependence $\tau _c(\lambda ^{-1})$ at
different $\alpha _0$ is shown on the Fig.\ref{fig:te}. It is seen
that at $\alpha _0>0$ the dependence $\tau _c(\lambda ^{-1})$ has
maximum, while at $\alpha _0<0$ it does not. Moreover, for
negative nonlinearity there is sharp lowering (but not to zero) of
$\tau _c$ at certain $\lambda ^{-1}$. This demonstrates that a
negative feedback almost destroys long-range order especially at
small dipoles concentrations. This behavior shows that at $\alpha
_0>0$ nonlinear effects produce positive feedback thus enhancing
the long-range order in the system, while $\alpha _0<0$ the
feedback is negative so that long-range order is inhibited but not
completely destroyed even for large negative $\alpha _0$, see
curve for $\alpha _0=-20$ on the Fig.\ref{fig:te}. Also, at
sufficiently large positive $\alpha _0$ we can achieve very large
enhancement of $T_c$ as compared to its mean field value (see
curve for $\alpha _0=3$ on the Fig.\ref{fig:te}). Since in our
model we suppose that the value of $T_{cmf}$ is equivalent to
$T_c$ for pure PbTiO$_3$ (the end member of compound), we conclude
that large positive $\alpha _0$ give substantial increase of $T_c$
as compared to the end members of ferroelectric compound.

\section{Comparison with the experiment. Discussion.}

In our model, the ferroelectric compounds are considered as an
ensemble of electric dipoles embedded in some virtual paraelectric
host, its nature we will discuss later. It can be supposed that in
the considered system (BiScO$_3$)$_{1-x}$(PbTiO$_3$)$_x$ electric
dipoles are originated from PbTiO$_3$, i.e. their number increase
with $x$ increasing. The nondipolar random field sources like
point charges, elastic dipoles etc. are also present in such
compositions due to the mixed valency of Bi and the difference in
charges and ionic radii of Bi$^{3+}$ and Pb$^{2+}$, Ti$^{4+}$ and
Sc$^{3+}$. These defects could easily be incorporated in the
consideration (see \cite{Glfarstef} for details), but we do not
consider their contribution here to have the minimal number of
adjustable parameters when fitting the theory with experiment.

\begin{figure}[th]
\vspace*{-5mm} \hspace*{-5mm} \centering{\
\includegraphics[width=8cm]{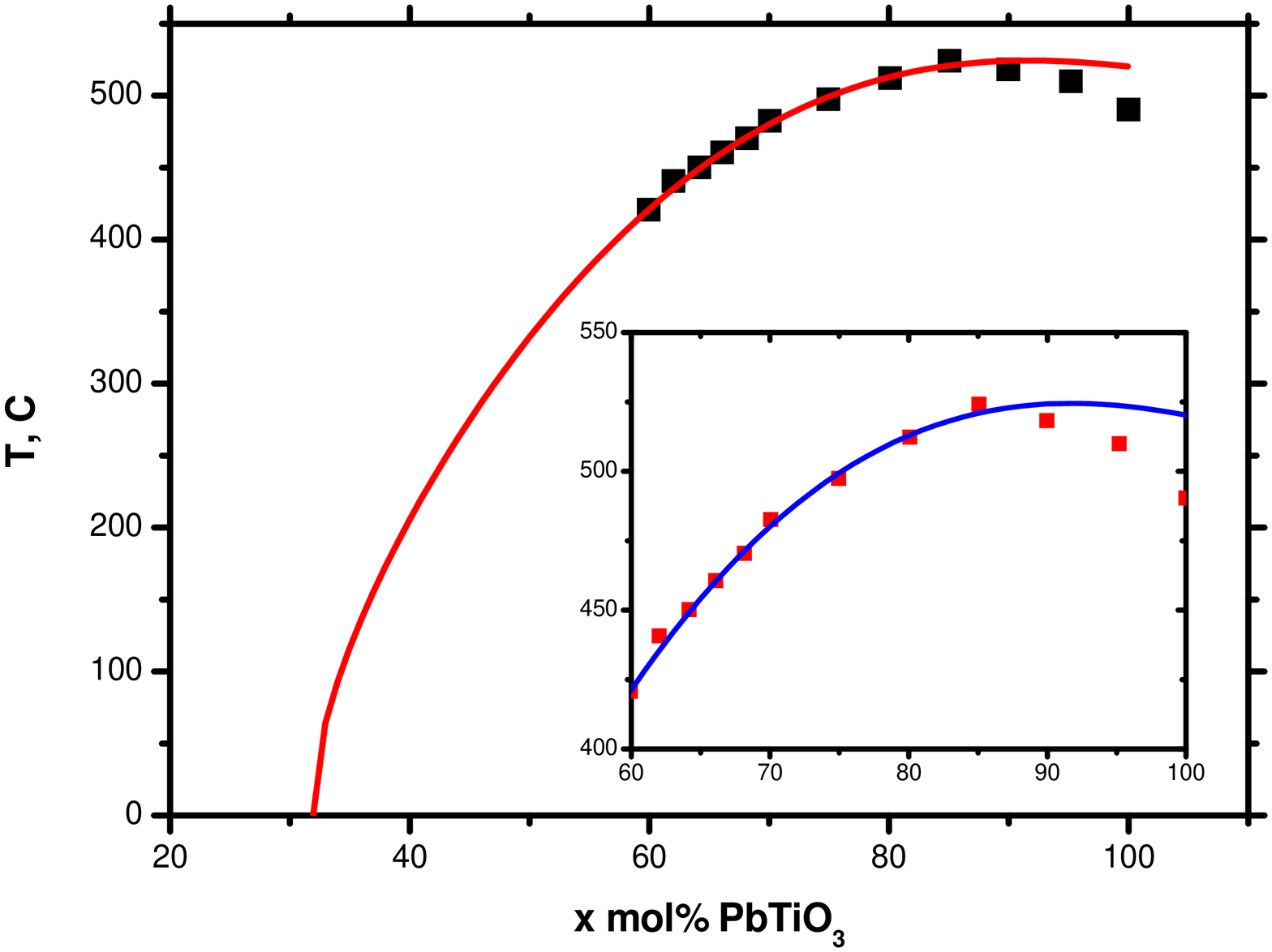}}
\caption{Comparison of our theory (line) with experiment
{\protect\cite{rb2}} (squares). Inset shows the fit in more
details.} \label{fig:ek}
\end{figure}

For quantitative description of experiment in above ferroelectric
compounds in the framework of our random field theory (in linear
or nonlinear approximation) we should have precise information
about the concentrations of electric dipoles and other random
fields sources as well as the parameters $\varepsilon _0$, $r_c$
etc. Unfortunately the available data are strongly restricted for
above compounds. That's why we recalculated our dimensionless
parameters (\ref{dim}) from the best fit to the experiment
\cite{rb2}. Namely, from the position and "amplitude" of maximum
of experimental curve $T_c(x)$ we determine our nonlinearity
coefficient $\alpha _0=0.81$ and coefficient of recalculation of
parameter $\lambda $ into $x$. Latter coefficient gives us value
$T_{cmf}=490^{\circ }$C and critical content of PbTiO$_3$
$x_{cr}=32\%$. The result of such fitting is shown on the
Fig.\ref{fig:ek}.

It is seen in Figure 2 that there is pretty good coincidence
between theory and experiment. This coincidence shows, that
physical mechanism of enhancement of ferroelectric phase
transition temperature beyond the end members in the ferroelectric
compounds is the enhancement of initial (i.e. that of end members)
long range ferroelectric order in them caused by positive feedback
generated by the nonlinear effects with positive coefficient. The
obtained value $T_{cmf}=490^{\circ }$C is very close to transition
temperature of PbTiO$_3$. This speaks in favour of statement that
PbTiO$_3$ paraelectric phase can be considered as a host lattice.

As we have mentioned above, the origin of nonlinear effects in
compound ferroelectrics is quite naturally related to its
intrinsic nonlinearities, reflected, for instance, in their
hysteresis loop. The physical origin of these nonlinearities may
be the nonlinear coupling of the ions in the unit cell as well as
the clusterization of dipoles in diluted ferroelectric compounds.
Each cluster has its own mesoscopic dipole moment (polarization of
cluster) and such clusters can interact between each other. If
interaction between such clusters is of positive sign, which may
be manifested in the almost rectangular shape of hysteresis loop,
we have the positive feedback with enhancement of initial (i.e.
that of end member) long-range order with positive nonlinear
coefficient. If the interaction is of negative sign, we have
negative feedback and inhibition of ferroelectric order. But such
negative feedback cannot destroy ferroelectricity completely,
probably except for the case of extremely small dipoles
concentration. So, to achieve high positive values of nonlinear
coefficients (which are necessary to strongly enhance $T_c$), we
should have (or prepare) the end members of ferroelectric compound
with "as much as possible rectangular" hysteresis loop. The
contribution of nonlinearities to dielectric response and
hysteresis loop in mixed perovskites PZT with impurities was
discussed in Ref. \cite{tan}. The essential role of nonlinearity
effect in explanation of phase diagram peculiarities in mixed
ferroelectrics PbZr$_{1-x}$Ti$_x$O$_3$, BaZr$_x$Ti$_{1-x}$O$_3$
and mixed systems of ferroelectric relaxors was shown recently in
Ref. \cite{GES} and Ref. \cite{GESH} respectively.

To make more precise prediction of which components in
ferroelectric compound to use to increase $T_c$ beyond end members
value, the measurements of correlation radius, Lorentz factor and
ions shifts in both end members and entire above ferroelectric
compounds are highly desirable.

It should be finally noted, that the theory outlined here permits
one to calculate not only $T_c$ for above substances but virtually
any equilibrium and/or nonequilibrium thermodynamic
characteristics such as static and dynamic dielectric
susceptibility, specific heat etc as a functions of PbTiO$_3$
content, temperature, electric field and other external
parameters.


\begin{thebibliography}{99}
\bibitem{rb1} R.E.Eitel, C.A.Randall, T.R.Shrout, W.Hackenberger,
S.E.Park, Jap. J. Appl. Physics {\bf 40} (1), 5999-6002 (2001).

\bibitem{rb2} R.E.Eitel, C.A.Randall, T.R.Shrout, S.E.Park,
Jap. J. Appl. Physics {\bf 141} (4A), 2099-2104 (2001).

\bibitem{rb3} C.A.Randall, R.E.Eitel, T.R.Shrout, D. Woodward,
I.M. Reaney, J. Appl. Phys. {\bf 93} (11), 9271-9274 (2003).

\bibitem{rb4} T.H.Song, R.E.Eitel, T.R.Shrout, C.A.Randall, W.Hackenberger,
Jap. J. Appl. Phys. {\bf 42} (8), 5181-5184 (2003).

\bibitem{rb5} J.Inigues, D.Vanderbilt, L.Bellaiche, \prb {\bf 67} (2)
Art. No. 224107 (2003).

\bibitem{rb6} Inaguma, A. Miyaguchi, M. Yoshida, T. Katsumata, Y.
Shimojo, R.P. Wang and T. Sekiya, J. Appl. Phys. {\bf 95} (11),
231-265 (2004).

\bibitem{GS1}  M.D. Glinchuk, V.A. Stephanovich, J. Phys. : Condens. Matter, {\bf
6}, 6317 - 6327 (1994).

\bibitem{GS2}  M.D.Glinchuk, V.A.Stephanovich, Ferroelectrics, {\bf 169},
281-291 (1995).

\bibitem{Koshe} J.Ya. Korenblit, E.F.Shender, Uspehi Phis. Nauk, {\bf 157},
267-307 (1989).

\bibitem{SG} V.A.Stephanovich European Phys. Journ. B, {\bf 18}, 17-21 (2000).

\bibitem{Glfarstef}  M.D.Glinchuk, R.Farhi, V.A.Stephanovich,
J.Phys.:Condens.Matter, {\bf 9} 10247 (1997).

\bibitem{SS} Yu. G. Semenov, V.A.Stephanovich, \prb {\bf 66} Art. No. 075202 (2002).

\bibitem{SS2} Yu. G. Semenov, V. A. Stephanovich, \prb {\bf 67} Art. No. 195203 (2003).

\bibitem{sto}  A.M.Stoneham, \rmp {\bf 41}, 82 (1969).

\bibitem{book}  M.D.Glinchuk, V.G.Grachev, M.F.Deigen, A.B.Roitcin, L.A.Syslin,
Electrical effects in radiospectroscopy. Moscow, Nauka, 1981 (in
Russian).

\bibitem{Glkond}  M.D.Glinchuk, I.V.Kondakova, Sol.St.Commun., {\bf 96}, 529 (1995).

\bibitem{ncr} We perform the asymptotic analysis at $\tau _c\to 0$
by changing the variable $y=x/\tau _c$ in the Eq. (\ref{tauc}) and
expanding in small parameter $\tau _c$. The analysis for the case
$\lambda \to \infty$ is performed by substitution $z^2=\pi \lambda
^2x^2/4$ and further expansion in small parameter $\lambda ^{-1}$.

\bibitem{tan} Qi Tan, D. Viehland, \prb {\bf 53}, 14103 (1996).

\bibitem{GES} M.D.Glinchuk, E.A.Eliseev, V.A.Stephanovich, Ferroelectrics
{\bf 254}, 27 (2001).

\bibitem{GESH} M.D.Glincuhk, E.A.Eliseev, V.A.Stephanovich, B.Hilczer,
Ferroelectrics {\bf 254}, 13 (2001).

\end{thebibliography}
\end{document}